\documentstyle[preprint,prl,aps]{revtex}
\title{Landau Ginzburg Theory and Nuclear Matter at Finite
Temperature\footnotemark}
\author{R. A. Ritchie$^{a,c}$,
H. G. Miller$^{a,b\rule{.05em}{.0em}}$\footnotemark\hspace{.1em} and
F. C. Khanna$^{a,c}$}
\address{$^{\mbox{\scriptsize a}}$
         Theoretical Physics Institute, Department of Physics,
         University of Alberta, Edmonton, Alberta, Canada T6G~2J1\\
         $^{\mbox{\scriptsize b}}$
         Department of Physics, University of Pretoria, Pretoria 
         0002, South Africa\\
         $^{\mbox{\scriptsize c}}$
         TRIUMF, 4004 Wesbrook Mall, Vancouver, British Columbia,
         Canada V6T~2A3}

\tightenlines
\newcommand{\be}{\begin{equation}}
\newcommand{\ee}{\end{equation}}
\newcommand{\ben}{\begin{eqnarray}}
\newcommand{\een}{\end{eqnarray}}

\begin{document}
\footnotetext[1]{We dedicate this paper to Prof.\ R. H. Lemmer on the 
occassion of his 65$^{\mbox{\scriptsize th}}$ birthday.}
\footnotetext[2]{Permanent Address}
\footnotetext{
 E-Mail:   \begin{tabular}[t]{l}     rritchie@phys.ualberta.ca\\
                 hmiller@scientia.up.ac.za\\
                 khanna@phys.ualberta.ca \end{tabular}}

\maketitle

\begin{abstract}
Based on recent studies of the temperature dependence of the energy 
and specific heat of liquid nuclear matter, a phase transition is 
suggested at a temperature $\sim .8$~MeV\@.  We apply Landau Ginzburg
theory to this transition and determine the behaviour of the energy and specific heat close to the critical temperature in the condensed phase.
\vspace{1em}
\end{abstract}

%\pacs{}
%\newpage
The existence of an energy gap in the spectrum of even-even nuclei
due to paired states of either protons or neutrons~\cite{BMP58} 
similar to that described by Bardeen, Cooper and Schrieffer (BCS) for
electrons in a superconductor~\cite{BCS57} has led to the suggestion
that nuclear matter should also exist in  a condensed phase for  some
range of temperatures~\cite{ES60}. The properties of this superfluid
phase in both nuclear and neutron matter have been studied in the BCS
approximation using a variety of phenomenological 
forces~\cite{BCSapprox}
%% G69,BPPR69,CC69,C69,O70,O71,YC71,CCY72,T72,AO85
as well as more realistic interactions~\cite{BCSreal}. Remarkably,
all calculations yield qualitatively similar results for $^{1}S_{0}$  
pairing, namely that neutron matter exists in a condensed phase for 
$k_{F}$ less than about \mbox{$1.3-1.5$~fm$^{-1}$}.  Recent
calculations, using the Paris potential~\cite{Letal80}, by the
Catania group~\cite{BCLL90} have shown that only slight deviations
occur in nuclear matter.  Such modifications,
which can be characterized by the use of a smaller nuclear effective
mass in the case of nuclear matter, are known to give rise to a
slight decrease in the gap, $\Delta $.  Although  such calculations
suggest that such a low temperature phase should exist in both
nuclear as well as neutron matter this has not been taken into
account in,  for example, astrophysical calculations since it is
thought that it may be masked by other instabilities~\cite{LR78}.

In field theoretic language BCS theory is considered as the
spontaneous symmetry breaking of phase symmetry.  The condensed
phase, $e.g.$ the superconducting phase, is characterized by an order
parameter, ($\Delta$), which is zero at the critical temperature,
$T_{c}$.  It has been established that the order parameter and the
critical temperature fulfill the following approximate
relationship~\cite{LP80ii}
\begin{equation} %\label{gaptc}
\frac{\Delta_0 }{T_c }  \approx 1.76
\end{equation}
where $\Delta_{0}$ is the value of the energy gap at $T=0$ and
here we have taken the Boltzmann constant $k_B=1$. In the normal
phase the order parameter is zero. Interestingly enough the same 
relationship has been found to hold in the aforementioned
calculations in nuclear matter~\cite{BCLL90}. Furthermore, it has
been pointed out that the same relation between $\Delta_0$ and $T_c$
also describes spontaneous symmetry breaking of chiral symmetry in
QCD  if $T_{c }$ is taken to be $2 f_{\pi}$~\cite{BCS85}, where
$f_{\pi}$ is the pion decay constant. In all of the aforementioned
cases the order parameter is obtained from a gap-like equation with
appropriate quasiparticle interactions.

Recent studies of nuclear matter have suggested that the origin of 
collective states may ultimately be linked to symmetry 
rearrangement~\cite{HKU97}. This leads to a BCS-like condensed phase, 
separated from the normal phase, which has an order parameter that 
goes to zero at the critical temperature. Calculations in finite 
nuclei at finite temperature suggest that this provides a reasonable 
description of the vanishing of the collective degrees of 
freedom~\cite{KMQ96}.

Recently  it has been demonstrated that the low temperature behavior 
of the specific heat of symmetric nuclear matter could be obtained 
from a finite temperature extension of the semi-empirical mass 
formula~\cite{NJDetal93}.  The temperature dependence of the 
coefficients in the semi-empirical mass formula~\cite{S61} was 
determined by fitting to the canonical ensemble average
of the excitation energy of over 300 nuclei for temperatures $T\leq 
4$~MeV, using experimental information of the energy spectra of 
nuclei in the mass region $22 \leq $A$ \leq 250$.  The volume term 
was then used to determine the temperature dependence of the energy 
per nucleon and specific heat of nuclear matter.  This displayed some 
rather interesting aspects:
A structure in both the energy and the specific heat was observed at 
temperatures between .5 and 1.3~MeV (the structure in the specific 
heat is of course more pronounced). Below this temperature the 
behaviour of the specific heat was quite different from that expected 
for a Fermi gas of free nucleons~\cite{NJDetal93}.  This is not 
unexpected as the low lying energy spectra of most nuclei are 
predominantly collective in nature.  Above 1.3~MeV, the specific heat 
was essentially linear in temperature as is the case for a
Fermi gas, but with the somewhat surprising feature that the slope 
coefficient was considerably larger than that suggested by the Fermi 
gas bulk level density parameter,
\begin{equation} a_v\approx \frac{1}{15} \frac{m^*}{m},
\end{equation}
where it was assumed $m^* = (0.7 - 1.2) m$ \cite{m*}.
We propose that the larger slope might in fact be quite reasonable,
based upon comparison with the case of liquid $^3$He.
It is well known that at low temperatures, normal liquid $^3$He may
be treated as a Fermi gas of quasi-particles. However, due to the
strength of the interactions in the liquid, constants such as the 
quasi-particle mass are not easily calculated, and rather are 
determined from experiment. For liquid $^3$He an effective mass of 
$m^*\sim 3m$ \cite{FW71} is obtained by measurement of the specific 
heat.  Following a similar procedure, noting the similarity between 
the free interactions of $^3$He and of nucleons, one anticipates that 
the normal liquid of nuclear matter would exhibit properties that are 
quite similar to those of normal liquid $^3$He. In particular, an 
effective mass of $m^*\sim 2m$ is required to fit the specific heat 
given in \cite{NJDetal93}. It is important to note that in the 
treatment of liquid nuclear matter, it is usually assumed that the 
liquid may be replaced by a gas of quasi-particles with the mass of 
the quasi-particles being equal to that of free nucleons. Then the 
effect of nucleon-nucleon interactions leads to a new effective 
mass of the quasi-particle to be $\sim 0.7-0.8$.  It is to be 
stressed that the procedure followed in the case of normal liquid 
$^3$He, namely to use the experimental data on specific heat to
deduce the effective mass, is more satisfying.

In this paper, we propose that there is a second order phase 
transition in liquid nuclear matter with a critical temperature $T_c$ 
and an order parameter $\eta$.  We apply Landau Ginzburg theory to 
determine the thermodynamic properties of the condensed phase close 
to $T_c$, from information about the normal phase.  We follow the 
procedure used for liquid helium to determine the normal phase, 
namely modelling the system as a Fermi gas of quasi-particles 
with an effective mass $m^*\sim 2 m$, determined from the specific 
heat.  We find that the behaviour of the energy per nucleon and 
specific heat across the phase transition with $T_c\sim .8$~MeV to be 
consistent with that shown in \cite{NJDetal93}.

Landau and Ginzburg have provided a simple theory of phase 
transitions which approximates the free energy in the region around
$T_{c }$ and is most useful in analyzing the thermodynamics in this 
region.  In particular, using only knowledge about the uncondensed 
phase one is able to make predictions about quantities in the 
condensed phase, such as specific heat, magnetic susceptibility and 
compressability.  Moreover, Landau Ginzburg theory can be derived 
from microscopic  calculations \cite{LP80ii}.

Following the Landau Ginzburg formulation it is necessary first to
determine an expression for the free energy per nucleon $f(T)$ in 
both phases.  In what follows the subscript 1 will refer to the lower 
temperature (condensed) phase, and 2 to the higher temperature 
(uncondensed) phase.  For the uncondensed phase, we take a quadratic 
form for the energy per nucleon which follows from a low temperature 
Fermi gas approximation of a normal Fermi liquid,
\begin{equation}
W_2(T)=a_2 + k_2 T^2, \label{eqW2}
\end{equation}
where $a_2$ and $k_2$ are constants.
From the relations for the specific heat per nucleon in terms of $W$ 
and the entropy per nucleon $s$,
\begin{equation}
c_V=    \frac{\partial W}{\partial T} 
   = T\,\frac{\partial s}{\partial T},
\end{equation}
we are able to deduce the entropy per nucleon in the uncondensed 
phase,
\begin{equation}
s_2(T)=C_2 + 2 k_2 T, \label{eqs2}
\end{equation}
where $C_2$ is an unknown integration constant which later cancels 
out of the calculation.  From eqs.~(\ref{eqW2})
and (\ref{eqs2}) the free energy per nucleon in the higher 
temperature phase is given by
\begin{equation}
f_2(T)=a_2 - C_2 T - k_2 T^2. \label{eqf2}
\end{equation}

To determine the free energy per nucleon in the condensed phase, we 
make use of the Landau expansion  \cite{LP80} for the free energy per 
nucleon in terms of an order parameter $\eta$, which goes to zero at 
the transition to the uncondensed phase. This order parameter is 
related to the presence of pairing expected at lower temperatures and 
vanishes with the pairing gap $\Delta$ at some critical temperature 
$T_c$.  The free energy per nucleon expansion to order $\eta^4$ is
\begin{equation}
f_1(T,\eta)=f_2 + A \eta^2 + B \eta^4.
\end{equation}
Here $A$ and $B$ are functions of temperature.
The order parameter is determined by requiring the condensed phase to 
be stable below $T_c$ (i.e.\ $f_1$ should be minimized w.r.t.\ $\eta$).
This leads to
\begin{equation}
f_1=f_2 -\frac{A^2}{4 B}.
\end{equation}
Further, since $A$ is of opposite sign in the condensed and 
uncondensed phases, whilst $B$ is strictly positive \cite{LP80}, the 
lowest order expansion of $A$ in $T-T_c$ can be parametrized as
\begin{equation}
A(T)= a (T-T_c)\; 2 \sqrt{B(T_c)}.
\end{equation}
Note especially that $a>0$ is an essential requirement following from
the phase diagram \cite{LP80}. Substituting for $A(T)$, the free 
energy per nucleon near $T_c$ is given by
\begin{equation}
f_1(T)= (a_2-a T_c^2) + (2aT_c-C_2)T -(k_2 + a) T^2, \label{eqf1}
\end{equation}
where $f_2$ is taken from eq.~(\ref{eqf2}).

From the free energy per nucleon given by eq.~(\ref{eqf1}),
we can now determine the energy per nucleon in the condensed
phase near $T_c$,
\begin{eqnarray}
W_1(T)&=& (a_2 - a T_c^2) + (a + k_2) T^2\\
                &=&  a_1 + k_1 T^2 .\label{eqW1}
\end{eqnarray}
Comparing this to the uncondensed phase (eq.~(\ref{eqW2})) we
note that the $T$ dependence is also quadratic, but has a larger 
coefficient. Thus the specific heat is discontinuous across the phase 
transition, and is necessarily larger ($k_1 > k_2$)  in the condensed 
phase.

We now compare the structure of $W_1(T)$ and $W_2(T)$ to what has 
been determined from the finite temperature extension of the 
semi-emperical mass formula \cite{NJDetal93}.  Before proceeding, it 
should be noted that the energy per nucleon in nuclear matter is 
obtained from the volume term of the binding energy for finite sized 
nuclei.  It may be anticipated that any sharp features (e.g.\ kink in 
the energy per nucleon and discontinuity in specific
heat $\Delta c_V$) will appear smoothed out. Thus whilst comparison 
is still possible at a qualitative level, it is difficult to obtain 
quantitative estimates for the critical temperature and the 
discontinuity in the specific heat.

In the region \mbox{($.5 - 1.3$~MeV)} the energy per nucleon from
\cite{NJDetal93} is observed to show a peak above the simple $T^2$ 
behavior.  This is in good agreement with what might be expected from 
a smoothed out downwards kink in $W$ at $T_c$,
which follows from eqs.~(\ref{eqW2}) and (\ref{eqW1}).  Furthermore,
the specific heat in \cite{NJDetal93} shows a sharp drop in the 
region ($.5-1.3$~MeV) which agrees well with a smoothed out
discontinuous drop ($=(k_1-k_2)T_c$).  It should be noted that the 
specific heat above 1.3~MeV is very nearly linear, supporting the use 
of a quadratic temperature dependence of $W_2$, and that the slope
below .5~MeV is greater than that above 1.3~MeV, which is in good 
agreement with our prediction that $k_1>k_2$.

If we treat the uncondensed phase as a Fermi gas of quasi-particles 
with effective mass $m^*\sim 2 m$, as suggested earlier based on the 
linear behaviour of the specific heat,  we can estimate this jump in 
specific heat at the transition to a condensed phase, where there is 
pairing with an associated energy gap $\Delta$. This is given by 
\cite{LP80ii},
\begin{equation}
\Delta c_V\approx 2.43 c_V
\end{equation}
where $c_V$ is the specific heat per nucleon in the uncondensed 
phase.  The uncondensed phase parameters for the energy per nucleon 
given in eqn~(\ref{eqW2}) have been fitted in \cite{NJDetal93}, 
giving $k_2=1/6.7$~MeV$^{-1}$ and $a_2=-16$~MeV\@. Using this value 
for $k_2$ and assuming $T_c\sim .8$~MeV we find 
$\Delta c_V\sim .6$~MeV, which is consistent with the behaviour of 
the specifc heat per nucleon given  in \cite{NJDetal93}.

At temperatures considerably higher than $T_c$,
the energy per particle given by eq.~(\ref{eqW2}) will become 
positive.  It is reasonable to identify this with a transition from a 
Fermi liquid to a Fermi gas, at temperature 
$T_{\mbox{\scriptsize LG}}$.
Using eq.~(\ref{eqW2}) with the fitted parameters  $a_2$ and $k_2$  
taken from \cite{NJDetal93}, we estimate this transition temperature 
to be at $T_{\mbox{\scriptsize LG}}\approx 10$~MeV\@. This compares 
favourably with $T_{\mbox{\scriptsize LG}}\approx 15-20$~MeV in  
\cite{HJAZMLZ8384} and $T_{\mbox{\scriptsize LG}}\approx 5$~MeV 
(finite nuclei)  \cite{JPetal95}.

In summary, following suggestions of a pairing transition
in nuclear matter, we have applied Landau Ginzburg theory to provide
estimates for the thermodynamical properties across such a phase 
transition.  For information on the uncondensed phase, we assume a 
quadratic temperature dependence for the energy per nucleon.
We find that the behaviour of the energy per nucleon and specific 
heat across the phase transition with $T_c\sim .8$~MeV is consistent 
with that observed from the fit of finite nuclei.  Following an 
analogy with liquid $^3$He, we suggest that the essentially linear 
temperature behaviour of the specific heat observed in 
\cite{NJDetal93} is indicative that nuclear matter in the 
uncondensed (normal Fermi liquid) phase may be considered as
a Fermi gas  of quasi-particles with an effective mass $m^*\sim 2m$.  
Such observation appears quite reasonable.

Lack of experimental data on nuclear matter at finite temperature 
makes further refinement of the model difficult.  Experimental 
determination of the thermodynamic properties in heavy ion collisions 
would be extremely helpful for understanding the properties of 
nuclear matter at finite temperature. Clearly the energy per nucleon 
obtained in~\cite{NJDetal93} at temperatures above 1.3~MeV is much 
stiffer than that of a  Fermi gas of free nucleons, which is often
used in many
astrophysical calculations~\cite{LR78} which in turn should 
affect neutrino production rates in stars.  As this is the major 
cooling mechanism in these objects it would be interesting to see 
precisely how important this deviation is.

%%
%% It would also be interesting to study what physical consequences 
%% follow from the presence of a super-conducting ground state 
%% and larger effective mass.  In particular this might have an 
%% observable effect on Astophysical models.  
%% --> what effects, perhaps on neutrino production ?
%%  Also, the presence of a superconducting
%% ground state could influence pair exchange between nuclei.
%% --> this has already been done

\acknowledgements
We thank Prof.\ J. A. Tuszynski for several very useful discussions 
about Landau Ginzburg theory. The research is supported in part by 
the Natural Science and Engineering Research Council of Canada. 
H.G.M. acknowledges the support of the Foundation for Research 
Development, South Africa.  H.G.M. expresses his gratitude to the 
Theoretical Physics Institute of the U. of Alberta for their warm 
hospitality.

\end{document}